\documentclass[%
 reprint,
superscriptaddress,
 amsmath,amssymb,
 aps,
pra,
floatfix,
]{revtex4-2}
\usepackage[english]{babel}
\usepackage{graphicx}
\usepackage{dcolumn}
\usepackage{bm}
\usepackage{mathdots}
\usepackage{color}
\usepackage{xcolor}
\usepackage{ulem}
\usepackage{physics}
\usepackage{amsmath}
\usepackage[T1]{fontenc}
\begin{document}

\preprint{APS/123-QED}

\title{Harnessing optical disorder for Bell inequalities violation}

\author{Baptiste Courme} \email[Corresponding author: ]{baptiste.courme@lkb.ens.fr}
\affiliation{%
Laboratoire Kastler Brossel, ENS-Universite PSL, CNRS, Sorbonne Universite, College de France, 24 rue Lhomond, 75005 Paris, France 
}%
\affiliation{%
Sorbonne Université, CNRS, Institut des NanoSciences de Paris, INSP, F-75005 Paris, France}%

\author{Malo Joly}
\affiliation{%
Laboratoire Kastler Brossel, ENS-Universite PSL, CNRS, Sorbonne Universite, College de France, 24 rue Lhomond, 75005 Paris, France 
}%
\author{Adrian Makowski}
\affiliation{%
Laboratoire Kastler Brossel, ENS-Universite PSL, CNRS, Sorbonne Universite, College de France, 24 rue Lhomond, 75005 Paris, France 
}%
\affiliation{Institute of Experimental Physics, Faculty of Physics, University of Warsaw, Poland}
 \author{Sylvain Gigan}
 \affiliation{%
Laboratoire Kastler Brossel, ENS-Universite PSL, CNRS, Sorbonne Universite, College de France, 24 rue Lhomond, 75005 Paris, France 
}%
\author{Hugo Defienne}
 \affiliation{Sorbonne Université, CNRS, Institut des NanoSciences de Paris, INSP, F-75005 Paris, France}

\date{\today}

\begin{abstract}

{Bell inequalities are a cornerstone of quantum physics. By carefully selecting measurement bases (typically polarization), their violation certifies quantum entanglement. Such measurements are disrupted by the presence of optical disorder in propagation paths, including polarization or spatial mode mixing in fibers and through free-space turbulence. Here, we demonstrate that disorder can instead be exploited as a resource to certify entanglement via a Bell inequality test. In our experiment, one photon of a polarization-entangled pair propagates through a commercial multimode fiber that scrambles spatial and polarization modes, producing a speckle pattern, while the other photon remains with the sender. By spatially resolving the speckle intensity pattern, we naturally access a large set of random and unknown polarization projections. We show that this set is statistically sufficient to violate a Bell inequality, thereby certifying entanglement without requiring active correction techniques. Our approach provides a fundamentally new way to test Bell inequalities, eliminating the need for an explicit choice of measurement basis, and offering a practical solution for entanglement certification in real-world quantum communication channels where disorder is unavoidable.}
\end{abstract}
\maketitle

\section{\label{sec:level1}Introduction}

{From the fundamental demonstrations of non-locality~\cite{einstein_can_1935,aspect_experimental_1982} to practical applications in imaging~\cite{defienne_advances_2024}, computation~\cite{jozsa_role_2003}, and communication~\cite{gisin_quantum_2007}, quantum entanglement lies at the heart of modern quantum technologies. 
A major challenge across all these applications is preserving the fragile quantum correlations that underpin entanglement during transmission through a real-world channel.
In particular, the presence of heterogeneous structures along the photon's propagation paths, such as atmospheric turbulence, scattering, or mode mixing in optical fibers, can severely degrade performance or even prevent the proper functioning of these systems.
This limitation is present for example in current quantum key distribution (QKD) systems involving satellite-to-earth links~\cite{yin_satellite-based_2017,liao_satellite--ground_2017} or those relying on fiber-optic networks~\cite{inagaki_entanglement_2013,wengerowsky_entanglement_2019,neumann_continuous_2022}.
\begin{figure*} [!ht]
    \centering
    \includegraphics[width = 1\textwidth]{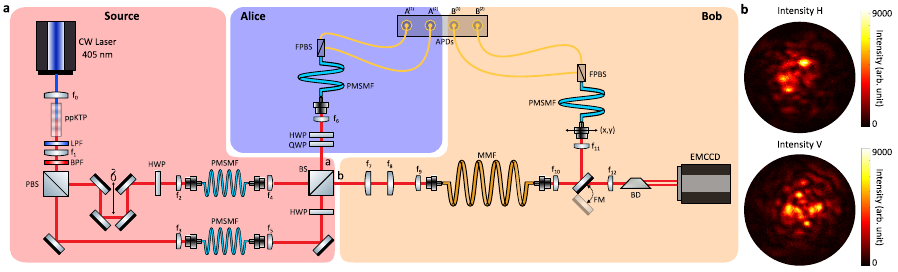}
   \caption{\textbf{Experimental setup.} 
\textbf{a.} \textit{Source:} A collimated continuous-wavelength (CW) laser at $405$~nm pumps a 1cm-long type-II periodically-poled crystal potassium titanyl phosphate (ppKTP) crystal. 
 {After the crystal, residual pump light is blocked with a low-pass filter (LPF), and the SPDC photons are spectrally selected using a $810\pm1$nm band-pass filter (BPF).} 
 Photons are rendered indistinguishable by adjusting the path delay $\delta$ {and are spatially filtered using a polarization-maintaining single-mode fiber (PMSMF). 
 Using half-wave plates (HWPs), the photon polarizations in each spatial path are set orthogonal to each other before being recombined at a beam splitter (BS) to create a polarization-entangled state. This arrangement is called a Shih-Alley configuration~\cite{shih_new_1988}. The output modes of the source are labeled $a$ and $b$.}
\textit{Alice:} On Alice’s side, photons pass through a quarter-wave plate (QWP) and a HWP before being coupled into a PMSMF connected to a fiber-based polarization beam splitter (FPBS). The two FPBS outputs are directed to avalanche photodiodes (APDs) labeled $A^{(1)}$ and $A^{(2)}$.\textit{ Bob:} On Bob’s side, photons are first injected into a $50$ cm long graded-index multimode fiber (MMF) with a $50\,\mu\text{m}$ core diameter.
At the output, photons are collected either with an electron-multiplied charge-coupled device (EMCCD) camera or with a PMSMF mounted on an XY translation stage. 
The fiber and the camera are positioned in planes conjugate to the fiber output facet. 
A calcite beam displacer (BD) produces two intensity images, one for each polarization, on the camera. 
The PMSMF is connected to a FPBS, whose outputs are coupled to two APDs labeled $B^{(1)}$ and $B^{(2)}$.
\textbf{b.} Intensity images for both polarization at the output of the MMF recorded with the EMCCD camera. All lenses $f_1$–$f_{12}$ are used either for coupling light into fibers or in telescope configurations: {$f_1=50$mm}; $f_2=10$mm; $f_3=10$mm; $f_4=25$mm; $f_5=25$mm; $f_6=25$mm; {$f_7=150$mm; $f_8=150$mm;} $f_9=5$mm; $f_{10}=10$mm; $f_{11}=25$mm; $f_{12}=400$mm.}
    \label{fig 2}
\end{figure*}

To mitigate these effects, numerous active correction techniques have been developed. For instance, adaptive optics approaches have been implemented to compensate for turbulence in real time, thereby extending the distance over which quantum-secure communication is possible~\cite{gruneisen_adaptive-optics-enabled_2021,scarfe_fast_2025,pugh_adaptive_2020,liu_single-end_2019}. Similarly, wavefront shaping techniques have enabled the distribution of entangled photons through multimode fibers~\cite{valencia_unscrambling_2020,valencia_large-scale_2025} and scattering layers~\cite{defienne_adaptive_2018,lib_real-time_2020,courme_manipulation_2023,courme_non-classical_2025,shekel_shaping_2024}. 
{However, these active methods allow only partial correction of the disorder, being limited both by the complexity of the medium and its temporal dynamics, which makes them difficult to use in real-world applications.}

Interestingly, since light scattering in most practical scenarios can be considered a linear process, the action of disorder is a unitary operation and can therefore be interpreted as an optical change of basis~\cite{kim_transmission_2015}. Even in the presence of losses, the random mode mixing process induced by optical disorder can be exploited for many applications, including speckle-based interferometers~\cite{chakrabarti_speckle-based_2015}, optical simulators~\cite{rafayelyan_large-scale_2020,wetzstein_inference_2020}, and the implementation of reconfigurable linear quantum circuits~\cite{defienne_two-photon_2016,wolterink_programmable_2016,leedumrongwatthanakun_programmable_2020,makowski_large_2024,goel_inverse_2024,joly_harnessing_2025}.
Here, we harness the optical disorder induced by a commercial multimode optical fiber to certify entanglement without the need for active correction methods. 
One photon from a polarization-entangled pair is sent through the fiber, which forms a speckle on the other side, i.e. it redistributes its intensity and polarization across more than $400$ spatial and polarization modes, each corresponding to a random rotation on the Poincaré sphere. 
By detecting coincidences between the fiber’s spatial output modes and the polarization of the photon retained by the sender, we demonstrate on a significant fraction of the  modes a violation of a Bell inequality, thereby certifying entanglement in the two-photon state. 
In our experiment, the fiber thus serves both as the propagation channel for transmitting entanglement and as a passive tool for its certification.} 

\section{Experiment}

The experimental setup is illustrated in Figure~\ref{fig 2}. A polarization-entangled two-photon state {with wavelength $810 \pm 1 \,\text{nm}$} is generated using a {Shih-Alley configuration~\cite{shih_new_1988}} (red-shaded area). When the optical delay $\delta \approx 0$, the state post-selected by a coincidence measurement between the detectors positioned in spatial modes $a$ and $b$ can be written as
\begin{equation}
\ket{\Psi} = \frac{1}{\sqrt{2}} \left [ a^\dagger_{aH} a^\dagger_{bV} - a^\dagger_{aV} a^\dagger_{bH} \right] \ket{0},
\label{Bell's state}
\end{equation}
where $a^\dagger_{mp}$ is the creation operator for a photon in spatial mode $m \in \{a,b\}$ with polarization $p \in \{H,V\}$ (see Methods for more details). 
The output modes $a$ and $b$ are connected to Alice’s arm (blue-shaded area) and Bob’s arm (yellow-shaded area), respectively.

On Alice's arm, the photon propagates through polarization-control optics consisting of a quarter-wave plate (QWP) and a half-wave plate (HWP) before being injected into a polarization-maintaining single-mode fiber (PMSMF). The fiber is connected to a fiber-based polarization beam splitter (FPBS), whose two outputs are coupled to two avalanche photodiodes (APD). By appropriately rotating the HWP and QWP, the photons polarization can be measured in any desired basis.
{On Bob’s arm, the photon is injected into a commercial graded-index multimode fiber (MMF) supporting approximately $400$ modes (i.e. $200$ spatial modes per polarization). At the fiber output, they are collected either with an electron-multiplying charge-coupled device (EMCCD) - for control purpose only - or with a PMSMF that can be scanned along the transverse spatial directions $x$ and $y$. Both the PMSMF and the EMCCD are positioned in optical planes conjugate to the fiber output facet. A Calcite beam displacer (BD) is placed in the optical path toward the camera to image the two output polarizations. The SMF, in turn, is connected to a FPBS, whose two outputs are coupled to two APDs.} 
\begin{figure*} [!ht]
    \centering
    \includegraphics[width = 1\textwidth]{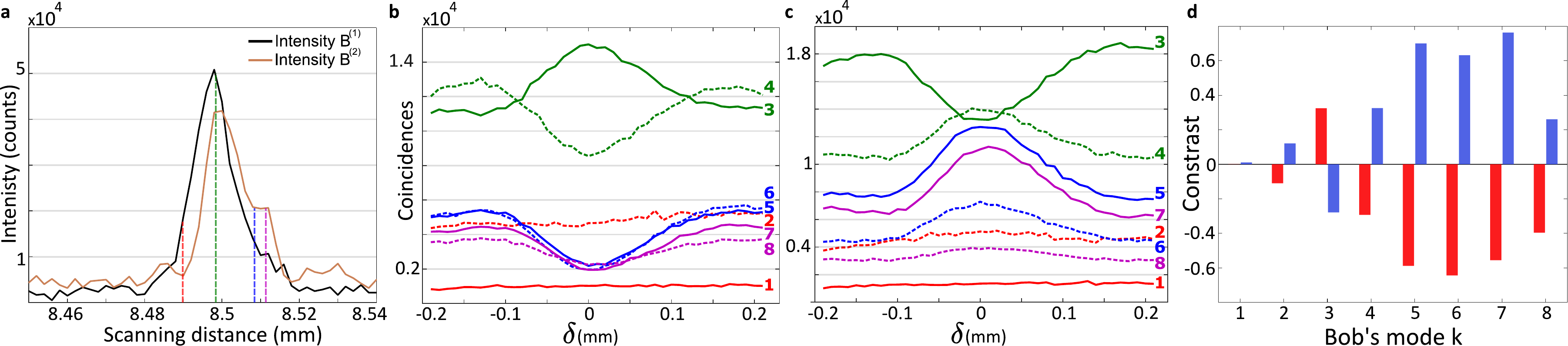}
\caption{\textbf{Complex projective measurements.} 
\textbf{a.}  Intensity measured at detectors $B^{(1)}$ (black curve) and $B^{(2)}$ (orange curve) while scanning the PMSMF along $x$ at a fixed $y$ position.
Four arbitrary positions are selected and indicated with dotted lines, forming a set of $8$ output modes in total, denoted $k \in [1,8]$.
\textbf{b.} Joint measurements between Alice (in setting $\hat{P}_D$) and Bob, corresponding to $\hat{P}_D \otimes \hat{P}_k$.
\textbf{c.} Joint measurements between Alice (in setting $\hat{P}_A$) and Bob, corresponding to $\hat{P}_A \otimes \hat{P}_k$. The colors of the solid and dashed lines refer to the spatial positions indicated by the matching colors in \textbf{(a)}. Coincidences were accumulated over 4 minutes for each PMSMF position.
\textbf{d.} Contrast values $C$ of the HOM-like curves as a function of Bob’s mode $k \in [1,8]$ and Alice’s projection $\hat{P}_D$ or $\hat{P}_A$. Red charts correspond to the Alice's projector setting $\hat{P}_D$ i.e. computed from the data in \textbf{(b)}. Blue charts correspond to the Alice's projector setting $\hat{P}_A$ i.e. computed from the data in \textbf{(c)}.}
    \label{fig 3}
\end{figure*}

{As shown in the intensity images in Figure~\ref{fig 2}b, light propagation through the MMF is scrambled, resulting in speckle patterns where the polarization varies from one speckle grain to another. Although complex, this mixing is deterministic and can be described by the transmission matrix $t$ of the MMF~\cite{popoff_measuring_2010,ploschner_seeing_2015}. Using this formalism, the input mode $b$ with polarization $p \in \{H, V\}$ is redistributed randomly over all spatial and polarization modes at the output of the MMF:
\begin{equation}
    a^\dagger_{bp} \rightarrow  \sum_{p'\in \{ H,V\}} \sum_{k=1}^M t^{p'p}_{kb} a^\dagger_{kp'},
\end{equation}
 where $M$ is the number of spatial modes of the MMF, $t^{p'p}_{kb}$ is the transmission matrix coefficient linking spatial input mode $b$ with polarization $p$ to spatial output mode $k$ with polarization $p'\in \{H, V\}$. 
The detection of a photon by one of Bob’s detectors, noted $B^{(i)}$ with $i\in \{1,2\}$, can thus be interpreted as a projective polarization measurement on the input spatial mode $b$. It is represented by a projector $\hat{P}_k^{(i)} = \ket{{\theta_{k}^{(i)} \phi_{k}^{(i)}}}\bra{{\theta_{k}^{(i)} \phi_{k}^{(i)}}}$, where
\begin{equation}
\left | {\theta_{k}^{(i)} \phi_{k}^{(i)}} \right \rangle = c_{k}^{(i)} \left[ \cos\left(\frac{\theta_{k}^{(i)}}{2}\right) \ket{H} + e^{i \phi_{k}^{(i)}} \sin\left(\frac{\theta_{k}^{(i)}}{2}\right) \ket{V} \right].
\label{projectionstate}
\end{equation}
As detailed in Methods, the parameters $c_{k}^{(i)}$, $\theta_{k}^{(i)}$, and $\phi_{k}^{(i)}$ depend on the transmission matrix coefficients $t^{HV}_{kb}$, $t^{HH}_{kb}$, $t^{VV}_{kb}$, and $t^{VH}_{kb}$, as well as on the specific Bob's detector $B^{(i)}$ considered.
Due to the complex multimode mixing in the MMF, the projectors $\hat{P}_k^{(1)}$ and $\hat{P}_k^{(2)}$ are uncorrelated~\cite{uppu_quantum_2016}. We therefore simplify the notations to $\hat{P}_k$, $c_{k}$, $\theta_{k}$ and $\phi_{k}$ throughout the remainer of the manuscript, where $k$ denotes a generic output mode defined by both the transverse spatial position $(x,y)$ of the PMSMF and Bob’s detector $\{ 1,2\}$.}

\section{Complex projective measurements}

{To highlight the complexity of Bob’s projections, we perform coincidence measurements between multiple fiber output spatial modes and the polarization state of Alice’s photon.
On Bob's side, we first scan the PMSMF along $x$ at fixed $y$ while recording single counts at detectors $B^{(1)}$ and $B^{(2)}$, as shown in Figure~\ref{fig 3}a. Then, four arbitrary positions are selected, forming a set of $8$ output modes $k \in [1,8]$.
On Alice's side, the HWP is set to $22.5^\circ$ and the QWP to $0^\circ$, so detections by $A^{(1)}$ and $A^{(2)}$ correspond to projections onto $\ket{D} = (\ket{H} + \ket{V})/\sqrt{2}$ and $\ket{A} = (\ket{H} - \ket{V})/\sqrt{2}$, respectively. The corresponding projectors are noted $\hat{P}_D$ and $\hat{P}_A$.

Coincidence rates between Alice and Bob detectors are then recorded for each selected position. 
Formally, each joint measurement corresponds to the action of one of the 16 projectors $\hat{P}_p \otimes \hat{P}_k$ on the input state $\ket{\Psi}$, with $p \in \{D, A\}$ and $k \in [1,8]$. 
While doing so, the delay $\delta$ of the Shih–Alley source is varied, yielding eight curves corresponding to Alice’s projection $\hat{P}_D$ (Fig.~\ref{fig 3}b), and another eight corresponding to her projection $\hat{P}_A$ (Fig.~\ref{fig 3}c). The shapes of these curves are similar to those observed in a Hong–Ou–Mandel (HOM) experiment~\cite{hong_measurement_1987}. 
In our case, setting $\delta = 0$ corresponds to the Shih–Alley source emitting an entangled state (Equation~\eqref{Bell's state}), whereas for $\delta \gg l_c$ (with $l_c \approx 0.1$ mm the photon coherence length), the state becomes separable (see Methods). The presence of `dips' or `bumps' in the HOM-like curves clearly demonstrates that the joint measurements performed by Alice and Bob are sensitive to the presence of entanglement in the input state. 

More quantitatively, we define the contrast of a HOM-like curve as $C = (R_{\delta=0} - R_{\delta \gg l_c})/{R_{\delta \gg l_c}}$, where $R_\delta$ denotes the measured coincidence rate at delay $\delta$. In this convention, positive contrasts correspond to bump-like features, whereas negative contrasts correspond to dips. As detailed in Methods, the contrast is theoretically related to Alice and Bob projections through the relation
\begin{equation}
  C = \epsilon \left| \sin \left({\theta_k}\right)\right| \cos \left( \phi_k \right),  
  \label{visibility}
\end{equation}
where $\epsilon = 1$ or $0$ depending on whether Alice performs $\hat{P}_D$ or $\hat{P}_A$, respectively, and $\theta_k$ and $\phi_k$ are the Poincaré sphere angles associated with Bob’s projection $\hat{P}_k$. 
Figure~\ref{fig 3}d shows the contrast of each HOM-like curve as a function of Bob’s mode $k$ and Alice’s projection. As predicted by Equation~\eqref{visibility}, the contrasts measured for Alice’s two projections are of opposite sign for a given mode $k$ of Bob. 
More importantly, strong variations of contrast are observed across Bob’s modes, with values ranging from –0.7 to 0.7. This demonstrates that $\theta_k$ and $\phi_k$ take very different values depending on the selected mode, highlighting the complexity of the projections implemented by the MMF.

\section{Clauser–Horne–Shimony–Holt (CHSH) violation}

This set of complex projections, enabled by the MMF mode mixing process, can be used as an efficient tool to certify entanglement through the violation of a Bell inequality. 
For that, we now select a total of $15$ transverse spatial positions on Bob’s side where the PMSMF can be placed, corresponding to a total of $N=30$ projectors $\hat{P}_k$ with $k \in [1,30]$. 
For each pair of projectors $(\hat{P}_k, \hat{P}_{k'})$, we define a polarization basis $\mathcal{B}_K$ formed by their associated polarization states, where the index $K=(k,k')$ uniquely labels the basis.
This configuration provides access to $N(N-1)/2 = 435$ measurement bases on Bob’s side.
In parallel, Alice performs measurements in two polarization bases, noted $\mathcal{A}$ and $\mathcal{A'}$. 
These bases are chosen randomly, with each corresponding to a different angular setting of Alice’s HWP and QWP.
To certify entanglement, we consider the $S$-parameter defined in Ref.~\cite{clauser_proposed_1969} and in Methods. It is calculated from $8$ joint measurements performed between the two bases on Alice’s side, $\mathcal{A}$ and $\mathcal{A'}$, and two bases on Bob’s side, $\mathcal{B}_K$ and $\mathcal{B}_{K'}$. In our experiment, this allows the calculation of a total of $(N(N-1)/2))^2 = 189225$ values of $S$.

The experimental statistical distribution of all $S$ values is shown in Figure~\ref{fig 4} for an entangled input state ($\delta = 0$, blue histogram) and for a separable state ($\delta \gg l_c$, orange histogram). Obviously, most $S$ values show no entanglement ($<2$) since Bob’s measurement basis is random, but a small yet significant fraction correspond to nearly optimal measurement bases. In the entangled case, the tail of the blue histogram shows that several values exceed $2$. In particular, a total of $2618$ ($2.2 \%$) values exceed this threshold, including {$654$} ($0.69\%$) by more than five standard deviations, confirming a statistically significant violation of the Clauser–Horne–Shimony–Holt (CHSH) inequality and thus certifying the presence of polarization entanglement in the input state. As expected, no $S$ values exceed $2$ for the case of a separable state at the input. {Simulations using random projections by Alice and Bob, presented in the supplementary document, are in excellent agreement with the experimental results.}
\begin{figure} [!ht]
    \centering
    \includegraphics[width = 0.45\textwidth]{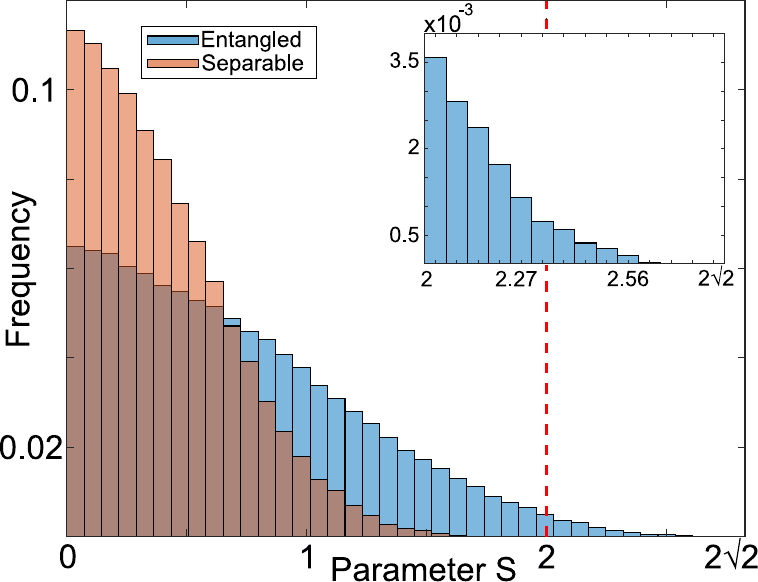}
 \caption{\textbf{Violation of the CHSH inequality.} Statistical distribution of the $189225$ values of the Bell parameter $S$ computed for an input entangled state ($\delta = 0$, blue histogram) and for an input separable state ($\delta \gg l_c$, orange histogram). The threshold $S = 2$ associated with the CHSH inequality is indicated by the red dotted line. Several values of $S$ exceed 2 in the entangled case, whereas none do in the separable case. In the non-entangled case, the maximum measured value is $S = 1.71 < 2$. In the entangled case, $2618$ values ($2.2\%$) exceed 2, including $654$ ($0.69\%$) that surpass the threshold by more than five standard deviations, demonstrating a statistically significant violation of the CHSH inequality. The inset shows the statistical distribution of the $S$ values above threshold. The total acquisition time is $4$ minutes per spatial position.}
    \label{fig 4} 
\end{figure}

\section{Discussion}

In this work, we demonstrate that the optical disorder of a commercial MMF can be harnessed to certify photon-pair entanglement. The large number of random projections available to Bob allows the computation of many $S$ values, some of which violate the CHSH inequality, thus providing a sufficient condition for entanglement certification.

A key advantage of our approach is its passive nature. Unlike adaptive optics or wavefront shaping~\cite{valencia_unscrambling_2020,valencia_large-scale_2025,defienne_adaptive_2018,lib_real-time_2020,courme_manipulation_2023,courme_non-classical_2025,shekel_shaping_2024}, it requires no pre-characterization and is not limited by correction performance.
For any medium, the probability of observing a violation depends on its disorder complexity: the stronger the disorder, the richer the diversity of projections, and the higher the probability.
In multimode fibers, this probability is thus determined by both the fiber’s intrinsic properties (e.g. type, core diameter, length) and the input excitation mode~\cite{ploschner_seeing_2015}.
The method also remains effective under dynamic conditions, provided that acquisition is faster than the decorrelation time, with temporal fluctuations of the disorder even supplying additional random projections.

In practice, the method reaches its full potential with parallel multimode detection at the output. Recent advances with SPAD arrays~\cite{ndagano_imaging_2020,courme_manipulation_2023} and time-stamping cameras~\cite{ianzano_fast_2020,courme_quantifying_2023,guitter_accidental_2025} make such measurements experimentally feasible. In our setup, a sensor resolving all $400$ output modes could generate over $2\times 10^{10}$ values of $S$, ensuring easy observation of a maximal violation. Because Alice’s measurement bases are randomly chosen, the approach can also be extended to scenarios where both photons propagate through disorder, yielding even larger sets of $S$ values - an important feature for QKD protocols with distant sources~\cite{ekert_quantum_1991}. Finally, since MMFs and scattering media couple all degrees of freedom~\cite{mosk_controlling_2012}, the method can be extended to spatial or spectral entanglement, with certification achieved via other inequalities~\cite{lib_experimental_2025}.

By exploiting rather than correcting disorder, our approach opens new avenues for robust quantum communication. In satellite-to-Earth links, current protocols require continuous polarization compensation to counter turbulence, scattering, and satellite motion~\cite{liao_satellite--ground_2017,bedington_progress_2017}. Our method could bypass these constraints by leveraging channel disorder or by collecting photons at ground stations via spatially resolved MMFs. Similarly, fiber-based quantum networks need active compensation at each relay to counter depolarization and polarization drift~\cite{hubel_high-fidelity_2007,xavier_experimental_2009,shi_fibre_2021}. Using MMFs in the channel or at detection removes this requirement, simplifying implementation and improving resilience to vibrations, mechanical stress, and temperature fluctuations. More broadly, treating disorder as a resource could enable scalable, robust quantum technologies in real-world conditions.

\section*{Methods}

{\noindent \textbf{Details on Equation~\eqref{projectionstate}.} The parameters $c_{k}^{(i)}$, $\theta_{k}^{(i)}$, and $\phi_{k}^{(i)}$ can be written in function of the transmission matrix coefficients as follow:
\begin{eqnarray}
    |{c}_{k}^{(i)}| &=& \sqrt{|t^{p'V}_{kb}|^2+|t^{p'H}_{kb}|^2} \\
    \arg \left({c}_{k}^{(i)}\right) &=& \arg \left(t^{p'H}_{kb} \right) \\
    \theta_{k}^{(i)} &=& 2 \,\text{atan} \left(\left | \frac{t^{p'V}_{kb}}{t^{p'H}_{kb}} \right |\right) \\
    \phi_{k}^{(i)} &=& \arg \left(t^{p'V}_{kb} \right) -\arg \left(t^{p'H}_{kb} \right)
\end{eqnarray}
where $p'=H$ if $i=1$ (detector $B^{(1)}$) and $p'=V$ if $i=2$ (detector $B^{(2)}$). \\
\\
\noindent \textbf{General expression for Alice and Bob's joint measurement.} We first define a general measurement performed by Alice through the projector $\hat{P}_{\theta\phi} = \ket{\theta\phi}\bra{\theta\phi}$, where:
\begin{equation}
\label{porjectAlicegene}
    \left | {\theta \phi} \right \rangle = \left[ \cos\left(\frac{\theta}{2}\right) \ket{H} + e^{i \phi} \sin\left(\frac{\theta}{2}\right) \ket{V} \right].
\end{equation}
For example, using this definition, the standard measurement settings used by Alice are written: $\hat{P}_H = \hat{P}_{00}$, $\hat{P}_V = \hat{P}_{\pi0}$, $\hat{P}_D = \hat{P}_{\pi/2\,0}$, and $\hat{P}_A = \hat{P}_{\pi/2\,\pi}$.
Then, we calculate the outcome of a general joint measurement by Alice and Bob on the input state. It can be expressed as:
\begin{eqnarray}
    & &\bra{\Psi}\hat{P}_{\theta \phi}\otimes \hat{P}_k \ket{\Psi} = \frac{1}{2} |c_k|^2  \bigg [ \left| \cos \left(\frac{\theta}{2}\right) \cos \left(\frac{\theta_k}{2}\right) \right|^2 \nonumber \\
    &+& \left| \sin \left(\frac{\theta}{2}\right) \sin \left(\frac{\theta_k}{2}\right) \right|^2 \\  
    &-& 2 \left| \cos \left(\frac{\theta}{2}\right) \cos \left(\frac{\theta_k}{2}\right) \sin \left(\frac{\theta}{2}\right) \sin \left(\frac{\theta_k}{2}\right)\right| \cos \left( \phi_k - \phi \right) \bigg] \nonumber.
     \label{measurement}
\end{eqnarray}
\\
\\
{\noindent \textbf{Details on the Shih-Alley source of polarization-entangled photons pairs.} This source has been introduced in Ref.~\cite{shih_new_1988}.} In our work, as detailed in Figure~\ref{fig 2}a, wavelength-degenerate photon pairs at $810$nm are produced inside a long periodically poled Potassium Titanyl Phosphate (ppKTP) crystal, cut for type-II phase matching. The pump beam is filtered out after the crystal using a long-pass filter (LPF). The orthogonally polarized photons are separated using a polarizaing beam splitter (PBS). The temporal delay between the two photons is controlled by adjusting the relative path length $\delta$ using a translation stage. 
After rotating the polarization of one photon with a HWP at $22.5^\circ$ so that both have the same vertical polarization, each photon is spatially filtered by coupling into a PMSMF.
At this stage, we obtain what is known as a source of indistinguishable photons, which can, for instance, be used to perform a Hong–Ou–Mandel (HOM) experiment by coupling the photons to a fiber beam splitter~\cite{hong_measurement_1987}. In doing so, we measure a maximum HOM-dip visibility of $V=93\%$, which indicates a high degree of indistinguishability between the photon pairs. Such a state can be written: 
\begin{equation}
    \ket{\phi} = a^\dagger_{cV} a^\dagger_{dV} \ket{0}, 
\end{equation}
where $c$ and $d$ denotes the spatial modes associated with the two PMSMFs.\\
Although the two photons are indistinguishable, the source does not produced a polarization-entangled state yet. To achieve this, one can use a Shih–Alley configuration~\cite{shih_new_1988}.
After rotating the polarization of one photon with a HWP at $22.5^\circ$ so that the polarizations are orthogonal, the photon pairs exiting the two PMSMFs are injected into the two input modes of a free-space beam splitter (BS). At the output of the BS, the state becomes:
\begin{equation}
\label{fullstate}
    \ket{\psi} = \frac{1}{2} \left[a^\dagger_{aH} a^\dagger_{aV} + i a^\dagger_{bH} a^\dagger_{bV} +a^\dagger_{aH} a^\dagger_{bV} - a^\dagger_{aV} a^\dagger_{bH} \right] \ket{0},
\end{equation}
where $a$ and $b$ are the two spatial output modes of the BS. 
Interestingly, when measuring coincidences between modes $a$ and $b$, only the terms $a^\dagger_{aH} a^\dagger_{bV}$ and $a^\dagger_{aV} a^\dagger_{bH}$ contribute. 
In this case, the state yields the same results as if a truly entangled state were used - an effect known as post-selection. 
Therefore, as long as modes $a$ and $b$ remain physically separated between the source and detectors, and coincidences are measured between them (as in our experiment), this source can be considered to generate the polarization-entangled two-photon state described in Equation~\eqref{Bell's state}.\\
To characterize the entangled state we produce, one can measure the parameter $S$ defined in Equation~\eqref{S parameter}. For this, Bob’s arm is replaced with one identical to Alice’s, consisting of a HWP and QWP to control polarization, a PMSMF coupled to a FPBS, which is connected to Bob’s two detectors, $B^{(1)}$ and $B^{(2)}$.
{By adjusting the HWP and QWP settings for Alice and Bob~\cite{clauser_proposed_1969}, we measure $S = 2.21 \pm 0.02 > 2$. Although not optimal, this value is sufficient to certify the presence of entanglement at the source.}
}\\
\\
{\noindent \textbf{Separable mixed state $\rho$.} When $\delta \gg l_c$, the photons no longer interfere coherently at the Shih–Alley source BS, preventing the generation of the post-selected entangled state. In this case, the state produced by the source is no longer the entangled state of Equation~\eqref{Bell's state}, but a mixed separable state $\rho$ given by:
\begin{equation}
\label{clacorrstate}
\rho = \frac{1}{{2}} \left[ a^\dagger_{aH} a^\dagger_{bV} |0\rangle \langle 0| a_{aH} a_{bV} - a^\dagger_{aV} a^\dagger_{bH} |0\rangle \langle 0| a_{aV} a_{bH} \right].
\end{equation}
In this configuration, a joint measurement by Alice and Bob using the operators $\hat{P}_{\theta\phi}$ and $\hat{P}_{k}$ (see Equations~\eqref{porjectAlicegene} and \eqref{projectionstate}) can be written as:
\begin{equation}
     \langle \rho \rangle = \frac{1}{2} |c_k|^2  \left [ \left| \cos \left(\frac{\theta}{2}\right) \cos \left(\frac{\theta_k}{2}\right) \right|^2 + \left| \sin \left(\frac{\theta}{2}\right) \sin \left(\frac{\theta_k}{2}\right) \right|^2 \right].
     \label{incoherentjointcorr}
\end{equation}}\\
\\
\noindent \textbf{Demonstration of Equation~\eqref{visibility}.} Combining the definition of contrast with Equations~\eqref{incoherentjointcorr} and~\eqref{measurement}, we obtain the analytical expression of $C$ as
\begin{equation}
  C = - 2 \frac{\left| \cos \left(\tfrac{\theta}{2}\right) \cos \left(\tfrac{\theta_k}{2}\right) \sin \left(\tfrac{\theta}{2}\right) \sin \left(\tfrac{\theta_k}{2}\right) \right| \cos \left( \phi_k - \phi \right)}{ \left| \cos \left(\frac{\theta}{2}\right) \cos \left(\frac{\theta_k}{2}\right) \right|^2 + \left| \sin \left(\frac{\theta}{2}\right) \sin \left(\frac{\theta_k}{2}\right) \right|^2 },  
\end{equation}
where $(\theta_k,\phi_k)$ are the angles associated with Bob’s projection $\hat{P}_k$, and $(\theta,\phi)$ are the angles associated with Alice’s projection $\hat{P}_{\theta \phi}$. Substituting $(\theta,\phi) = (\pi/2,0)$ and $(\pi/2,\pi)$ to implement $\hat{P}_D$ and $\hat{P}_A$, respectively, one recovers Equation~\eqref{visibility}.
 \\
\\
\noindent \textbf{Additional details on the experimental setup. }
The crystal is pumped by a $405$nm single-mode continuous-wave laser (DLproHP, Toptica) in a single spatial mode configuration. The 10-mm ppKTP crystal is placed in an oven, with its temperature externally controlled to maximize the number of generated photon pairs. The BPF in centered at $810$nm and has a width of $1$nm. The MMF used in the setup is the Thorlabs GIF50C ($50\pm2.5 \mu m$ core diameter, $55.3\pm0.1$cm length,$NA = 0.2$). Detection is performed using APDs with an efficiency of approximately $50\%$ (Excelitas), and single-photon time tags are recorded using a Swabian Time Tagger. The temporal coincidence window used for the measurement was set to $2.5$ns.\\
\\
{\noindent \textbf{CHSH measurement and $S$-parameter.} To certify the presence of entanglement, we perform a CHSH measurement~\cite{clauser_proposed_1969}. The $S$-parameter is evaluated from experimental measurements on the state to be characterized and is formally defined as:
\begin{eqnarray}
S(\mathcal{A},\mathcal{A'},\mathcal{B}_k,\mathcal{B}_{K'})&=&  | E(\mathcal{A},\mathcal{B}_K)+
E(\mathcal{A'},\mathcal{B}_K) \nonumber \\
&+& E(\mathcal{A},\mathcal{B}_{K'})-E(\mathcal{A'},\mathcal{B}_{K'})|,
    \label{S parameter}
\end{eqnarray}
where $\mathcal{A}$, $\mathcal{A'}$, $\mathcal{B}_K$, and $\mathcal{B}_{K'}$ are polarization bases used by Alice and Bob to perform their joint measurements, and $E$ is called the correlation value between two bases. To define it in general, we consider two bases $\mathcal{A} = \{\ket{A_1}, \ket{A_2}\}$ and $\mathcal{B} = \{\ket{B_1}, \ket{B_2}\}$, with associated projectors $\hat{P}_{A_1} = \ket{A_1}\bra{A_1}$, $\hat{P}_{A_2} = \ket{A_2}\bra{A_2}$, $\hat{P}_{B_1} = \ket{B_1}\bra{B_1}$, and $\hat{P}_{B_2} = \ket{B_2}\bra{B_2}$. The correlation parameter $E$ is then given by:
\begin{equation}
 E(\mathcal{A},\mathcal{B}) = \frac{R_{A_1B_1} - R_{A_1B_2} - R_{A_2B_1} + R_{A_2B_2}}{R_{A_1B_1} + R_{A_1B_2} + R_{A_2B_1} + R_{A_2B_2}},   
\end{equation}
where $R_{A_iB_j} = \bra{\Psi} \hat{P}_{A_i} \otimes \hat{P}_{B_j} \ket{\Psi}$ for $(i,j) \in \{1,2\}$ is the result of the joint measurement between Alice and Bob on the projections $\hat{P}_{A_i}$ and $\hat{P}_{B_j}$.\\
In practice, to demonstrate entanglement in the state $\ket{\Psi}$, it is sufficient to find two pairs of bases $\mathcal{A},\mathcal{A'}$ for Alice and $\mathcal{B},\mathcal{B'}$ for Bob such that measurements in these bases yield a value of $S > 2$. Importantly, failing to observe $S > 2$ does not imply the state is separable; it only means no conclusion can be drawn. In a typical Bell experiment, the measurement settings of Alice and Bob (HWPs and QWPs) are chosen optimally to maximize the violation, with a theoretical maximum of $S = 2\sqrt{2}$~\cite{clauser_proposed_1969}.
\\
\\
{\noindent \textbf{Measurement uncertainties.} Each measurement of an $S$ value is associated with a standard deviation. To determine it, we first estimate the standard deviation of each joint measurement between Alice and Bob by assuming that the counts follow a Poisson distribution i.e. the error corresponding to the number of counts detected
in a certain time interval is equal to the square root of that value. The corresponding uncertainty is then propagated to the $S$ values, as detailed in Ref.~\cite{sanz_undergraduate_2024}.}

\section*{Data availability}
Data are available from the corresponding author upon request.

\section*{Acknowledgments}

H.D. acknowledges funding from the ERC Starting Grant (No. SQIMIC-101039375) and the ANR (No. ANR-24-CE97-0001 and ANR-23-CE47-0014). AM acknowledges National Science Centre, Poland (2023/49/N/ST7/04195). S.G. acknowledges funding from the ANR (No. ANR-21-CE47-0021) and is supported by Institut Universitaire de France.

\section*{Author Contributions}
{B.C. analyzed the data, designed and performed the experiments, with support from M.L. and A.M. H.D. and S.G. conceived the original ideal and supervised the project. All authors discussed the results and contributed to the manuscript.}

\bibliography{references}

\newpage

\clearpage

\onecolumngrid

\section*{Supplementary document}

To validate the results presented in Figure 3 of the manuscript, we compared them with simulation results. To this end, we proceeded step by step:
\begin{enumerate}
    \item Alice randomly selects two projection bases, $\mathcal{A}$ and $\mathcal{A}'$, by randomly choosing angle pairs for her HWP and QWP. Each basis comprises two vectors with opposite orientations on the Poincaré sphere (i.e. $\ket{\theta \phi}$ and $\ket{-\theta -\phi}$), as shown in Figure~\ref{fig SM1}a.
    \item Bob chooses $N = 30$ vectors $\ket{\theta_k \phi_k}$ with angles $\theta_k$ and $\phi_k$ randomly distributed over $[0, 2\pi]$. They are represented in the Poincaré sphere in Figure~\ref{fig SM1}b.
    \item From all of Bob’s random projections, we construct $N(N-1)/2 = 435$ distinct bases, denoted $\mathcal{B}_K$ with $K \in [1,435]$.
    \item Using Equations (10) and (14), we compute all joint measurements between the vectors of Alice’s two bases, $\mathcal{A}$ and $\mathcal{A}'$, and those of any pair of Bob’s bases, denoted $\mathcal{B}_K$ and $\mathcal{B}_{K'}$. We also consider the intermediate case of photon pairs with a partial visibility of $V=93\%$, which corresponds to our experimental conditions. In this case, one can show that a joint probability measurement between Alice and Bob using the projectors $\hat{P}_{\theta \phi}$ and $\hat{P}_{k}$ can be written as:
\begin{eqnarray}
 \langle \rho \rangle &=& \frac{1}{2} |c_k|^2  \bigg [ \left| \cos \left(\frac{\theta}{2}\right) \cos \left(\frac{\theta_k}{2}\right) \right|^2 \nonumber 
+\left| \sin \left(\frac{\theta}{2}\right) \sin \left(\frac{\theta_k}{2}\right) \right|^2 \\  
&-& 2 V \left| \cos \left(\frac{\theta}{2}\right) \cos \left(\frac{\theta_k}{2}\right) \sin \left(\frac{\theta}{2}\right) \sin \left(\frac{\theta_k}{2}\right)\right| \cos \left( \phi_k - \phi \right) \bigg]. 
 \label{measurementgeneral}
\end{eqnarray}
Here, we therefore simulate the measurements between Alice and Bob for three different input states: a maximally entangled state ($V=1$), a mixed separable state ($V=0$), and a partially entangled state corresponding to our experimental conditions ($V=0.93$).
\item Using Equations (16) and (17), we compute the $(N(N-1)/2)^2=189225$ values of $S(\mathcal{A},\mathcal{A'},\mathcal{B}_K,\mathcal{B}_{K'})$. 
\end{enumerate}
We repeat this simulation for $100$ randomly chosen bases for Alice (i.e. two sets of $100$ random rotation angles for Alice’s HWP and QWP). The averaged distributions of the $S$-values are shown in Figure~\ref{fig SM1}c for the cases $V=1$ (black line), $V=0$ (orange histogram), and $V=0.93$ (blue histogram). The shapes of these curves are in very good agreement with the experimental results shown in Figure 3 of the manuscript.

\begin{figure} [!ht]
    \centering
    \includegraphics[width = 1\textwidth]{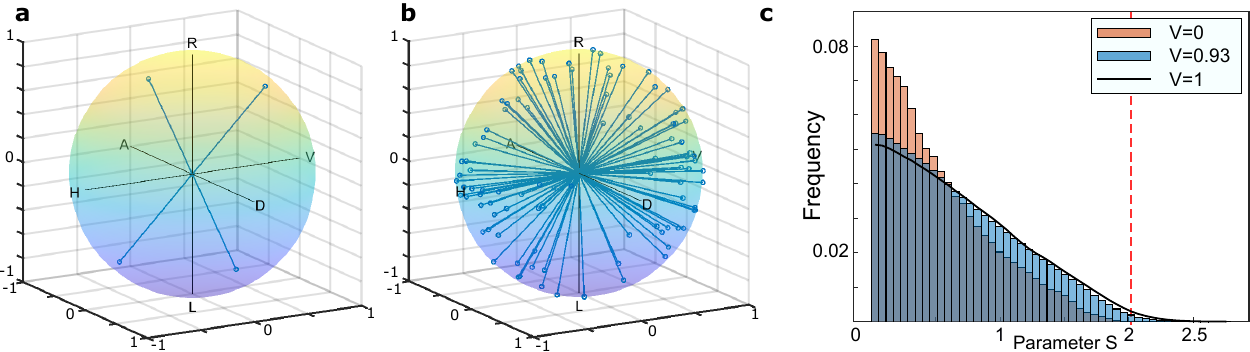}
 \caption{\textbf{Violation of the CHSH inequality with simulated random measurements on Alice’s and Bob’s sides.} \textbf{a.} Example of four simulated projections of Alice, obtained by random rotations of her HWP and QWP, represented on the Poincaré sphere. They form two bases, $\mathcal{A}$ and $\mathcal{A'}$.
\textbf{b.} $N=30$ simulated random projections for Bob, represented on the Poincaré sphere. \textbf{c.} Statistical distribution of the values of $S$ for a mixed separable state $V=0$ (orange histogram), a maximally entangled state $V=1$ (black line), and a partially entangled state with visibility $V=0.93$, as in our experiment (blue histogram).}
    \label{fig SM1} 
\end{figure}

\end{document}